\documentclass[a4paper]{jpconf}
\usepackage{graphicx}
\usepackage{amssymb}
\usepackage{amsmath}

\DeclareMathOperator{\Div}{div}

\begin{document}
  \title{The spin evolution of the pulsars with non-rigid core}
  
  \author{D. P. Barsukov$^{1,2}$, O. A. Goglichidze$^{1,*}$, A. I. Tsygan$^{1}$}
  
  \address{$^1$Ioffe Physical-Technical Institute of the Russian Academy of Sciences, Saint-Petersburg, Russian Federation \\
      $^2$Saint-Petersburg State Polytechnical University, Saint-Petersburg, Russian Federation}
      
  \ead{$^*$goglichidze@gmail.com}    
  \begin{abstract}
    We formulate a model of pulsar spin evolution (braking, inclination angle 
    evolution and radiative precession) taking into account the non-rigidity 
    of neutron star rotation.
    We discuss two simple limiting cases of this model and show that the evolution of the
    inclination angle substantially depends on the model of crust-core interaction.
    The non-rigidity of core rotation accelerates the inclination angle evolution and makes all pulsars evolve to the orthogonal state. The size of the effect depends on the amount of differentially rotating matter and mechanism of its interaction with the rest of the star. Since the rapid evolution of the inclination angle apparently contradicts 
    the observational data, our results may be used as an additional test for the theories of the cores 
    of neutron stars.
 \end{abstract}
  
  \section{Introduction}
    
  A rotating magnetized neutron star, if it has perfectly spherical shape, can be characterized by two vectors: angular velocity vector $\vec{\Omega}$ and magnetic moment vector $\vec{m}$. During the life of a neutron star the magnitudes of these vectors as well as the inclination angle $\chi$ formed by these vectors evolve in time. This evolution is caused by electromagnetic torque acting on rotating magnetized star.  A neutron star is surrounded by a large magnetosphere.
    Strictly speaking, the electromagnetic torque can be calculated only by using a self-consistent theory of this magnetosphere which despite the achieved progress is far from complete at present \cite{Spitkovsky2008}.

    The problem becomes even more complicated if one wishes to take into account the internal structure of neutron star. Neutron stars are not perfectly rigid. They contain a liquid core.     
    There are two main mechanisms of crust-core interaction: coupling through the magnetic field and viscosity \cite{Easson1979}.
    The first mechanism is much more effective and, if it works, then the protons, electrons and normal neutrons in the core can be considered  as rigidly rotating with the crust.
    In contrast, the superfluid neutrons which are believed to be present in neutron star's core \cite{YakovlevLevenfishShibanov1999} are decoupled from the rest of the matter and interact with it only by weak mutual friction force (vortex-mediated interaction) \cite{HallVinen1956}. 

    The configurations when the magnetic field does not penetrate the core have been discussed in the literature \cite{PonsMirallesGeppert2009,GourgouliatosEtAl2013}. The expulsion of magnetic field can be caused, for example, by type-I superconductivity of core protons.
    Despite the fact that the type-II proton superconductivity is more likely from the point of view of microscopic calculations (see, however, \cite{BuckleyMetlitskiZhitnitsky2004a}), its implications for the  neutron star rotation dynamics are not so clear. The coexistence of proton magnetic fluxoids with neutron superfluidity seems to be inconsistent with the observed long periods of precession in some neutron stars \cite{Link2006}.    
    
    In the present paper we consider in detail the two limiting cases of strong and weak coupling between 
    the crust and the core of a neutron star, which correspond to the two physical cases discussed above. 
    
  \section{Basic Assumptions}  
    We will treat a neutron star as a spherical rigid shell containing a liquid core in a spherical cavity. 
    We will denote the inner radius of the crust by $R_c$ and the outer one by $R_{ns}$. The crust rotates with angular velocity $\vec{\Omega}$. It feels the action of an external (electromagnetic) torque $\vec{K}$ 
    which is assumed to be slowly varying in the reference frame co-rotating with the crust. The core acts on the crust with a torque $\vec{N}$. Thus, the motion of the crust can be described as 
    \begin{equation}
      \label{eq:crust_eq}
      I_{crust}\dot{\vec{\Omega}} = \vec{K}+\vec{N},
    \end{equation}
    where $I_{crust}$ is the crust moment of inertia. Here and below we use the notation $\dot{\vec{\Omega}} = d_t\vec{\Omega}$.

    Let us for clarity consider the simplest possible system of hydrodynamical equations written down in the frame of reference co-rotating with the crust:
    \begin{equation}
      \label{eq:Euler}
      \partial_t\vec{u}+2[\vec{\Omega}\times\vec{u}]+(\vec{u}\cdot\vec\nabla)\vec{u}+\vec\nabla P/\rho+\vec\nabla\Phi= -[\dot{\vec{\Omega}}\times\vec{r}]+\vec{f},
    \end{equation}
    \begin{equation}
      \label{eq:cont}
      \partial_t\rho+\Div(\rho\vec{u})=0,     
    \end{equation}
    \begin{equation}
      \Phi(\vec{r}) = -G\int\frac{\rho(\vec{r}')}{|\vec{r}-\vec{r}'|}d^3r'-\frac{1}{2}[\Omega\times\vec{r}]^2. \label{eq:grav_pot}      
    \end{equation}
    Here, $\vec{f}$ is the sum of forces (per unit mass) acting on the fluid, $2[\vec{\Omega}\times\vec{u}]$ is the Coriolis force, $-[\dot{\Omega}\times\vec{r}]$ is the effective force caused by non-uniformity of neutron star rotation.

    For a neutron star it is a good approximation to assume 
    that the temperature is constant
    inside the star core \cite{GnedinYakovlevPotekhin2001}. It allows us to write the pressure gradient as
    \begin{equation}
      \label{eq:nabla_P}
      \vec\nabla P   = \sum_\alpha \rho_\alpha\vec\nabla\mu_\alpha
    \end{equation}
    where $\mu_\alpha$ is the chemical potential per unit mass of $\alpha$-th constituent, $\rho_\alpha$ is the $\alpha$-th constituent's mass density.
    
    Let us suppose that the flows in the core are produced only by $\dot{\vec{\Omega}}$. In the case of uniform rotation the star is in hydrostatical equilibrium:
    \begin{equation}
      \label{eq:HD_eql}
      \vec\nabla P^{(0)} + \rho^{(0)}\vec\nabla\Phi^{(0)} = 0.
    \end{equation}    
    Here and below the superscript $(0)$ denotes the hydrostatical values of variables.
    
    Since $\epsilon=\dot{\Omega}/\Omega^2\ll1$ one can consider $\vec{u}$ as a small perturbation to the rigid-body rotation $[\vec{\Omega}\times\vec{r}]$ produced by $\dot{\vec{\Omega}}$. All variables can be expanded in a power series in $\epsilon$:    
    \begin{align}
      \label{eq:eps_expan}
       {\rho}_\alpha =  {\rho}_\alpha^{(0)}+ {\rho}_\alpha^{(1)}+O(\epsilon^2), \ \ 
       {\Phi} =  {\Phi}^{(0)}+ {\Phi}^{(1)}+O(\epsilon^2), \ \ 
       {\mu}_\alpha =  {\mu}_\alpha^{(0)}+{\mu}_\alpha^{(1)}+O(\epsilon^2)
    \end{align}
    and the terms quadratic in $\epsilon$ can be neglected.   
    We also assume that there is some dissipation mechanism in the system that is characterized by a
time-scale $\tau_{dis}$ which is much smaller than the time-scale of rotational evolution $\tau_{re}=\Omega/\dot{\Omega}$ of the star. If so, the time-derivatives in Eqs. \eqref{eq:Euler} and \eqref{eq:cont} become negligibly small for the neutron stars with ages much larger than $\tau_{dis}$. Then
our  equations take the form
    \begin{equation}
      2[\vec{\Omega}\times\vec{u}]+\sum_\alpha \frac{\rho_\alpha^{(0)}}{\rho^{(0)}}\vec\nabla\tilde{\mu}_\alpha^{(1)}= -[\dot{\vec{\Omega}}\times\vec{r}]+\vec{f}^{(1)},
    \end{equation}
    \begin{equation}
      \Div\left(\rho^{(0)}\vec{u}\right)=0,    
    \end{equation}
    where we introduced $\tilde{\mu}_\alpha=\mu_\alpha+\Phi$. Two observations can be made. 
    First, neglecting the temperature term in $\vec\nabla P$, one is  able to exclude Eq. \eqref{eq:grav_pot} from consideration (if there is no need to calculate $\rho^{(1)}$, $\mu_\alpha^{(1)}$ or $\Phi^{(1)}$).
    Second, for neutron stars with ages much larger than $\tau_{dis}$ the internal flow ``forgets'' the initial conditions. At each time instance the velocity field is determined only by instantaneous values of $\vec{\Omega}$ and $\dot{\vec{\Omega}}$ and depends on time only through these two vectors (quasistationary approximation). 
    
    \section{Rotational dynamics}
    After the quasistationary approximation becomes valid the interaction torque can be written 
    in the following form
    \begin{align}
      \label{eq:N_torque}
       \vec{N} = - S_1 I_{core}\vec{e}_\Omega(\vec{e}_\Omega\cdot\dot{\vec{\Omega}})-S_2 I_{core}\vec{e}_\Omega\times[\vec{e}_\Omega\times\dot{\vec{\Omega}}]+S_3 I_{core}[\vec{e}_\Omega\times\dot{\vec{\Omega}}],
    \end{align}    
    where $I_{core}$ is the core moment of inertia, $S_1$, $S_2$, $S_3$ are some dimensionless coefficients and $\vec{e}_\Omega=\vec{\Omega}/\Omega$.
    It can be shown that $S_1 \approx 1$. The coefficients $S_2$ and $S_3$ are determined by 
    a specific crust-core interaction mechanism.
    
    The external torque can be written in the following form
    \begin{equation}
      \label{eq:K_torque}
       \vec{K} = K_0\left(\tilde{k}_\Omega\vec{e}_\Omega+\tilde{k}_m\vec{e}_m+\tilde{k}_\perp[\vec{e}_\Omega \times \vec{e}_m]\right),
    \end{equation}
    where  $\vec{e}_m =\vec{m}/m$,
    $\tilde{k}_\Omega$, $\tilde{k}_m$, $\tilde{k}_\perp$ are some dimensionless functions which depend on relative orientation between $\vec{\Omega}$ and $\vec{m}$ (and other possible variables specific for particular magnetosphere model). They are defined in such a manner that 
    $K_0$ equals  the value of the torque acting on aligned ($\chi=0$) pulsar.

    Substituting \eqref{eq:K_torque} and \eqref{eq:N_torque} into \eqref{eq:crust_eq} and solving the equation for $\dot{\vec{\Omega}}$ we obtain three scalar equations describing pulsar braking, inclination angle evolution and torque-driven precession 
    \begin{equation}
      \label{eq:omega_eq}
      \dot{\Omega} = \frac{K_0}{I_{tot}}\left(\tilde{k}_\Omega+\tilde{k}_m\cos\chi\right),
    \end{equation}
    \begin{equation}
      \label{eq:chi_eq}
      \dot{\chi} = -\frac{K_0}{\Omega}\frac{(I_{crust}-S_2 I_{core})\tilde{k}_m -S_3I_{core} \tilde{k}_\perp}{(I_{crust}-S_2 I_{core})^2+S_3^2 I_{core}^2}\sin\chi,
    \end{equation}
    \begin{equation}
      \label{eq:phi_eq}
      \dot{\varphi}_\Omega = -\frac{K_0}{\Omega}\frac{(I_{crust}-S_2 I_{core})\tilde{k}_\perp+S_3I_{core} \tilde{k}_m}{(I_{crust}-S_2 I_{core})^2+S_3^2 I_{core}^2},
    \end{equation}
    where $I_{tot}=I_{crust}+I_{core}$. Note that Eq.~\eqref{eq:omega_eq} has the same form as for perfectly rigid star. Roughly speaking, it is a consequence of the energy conservation law. 

    Eqs. \eqref{eq:chi_eq} and \eqref{eq:phi_eq} differ from analogous equations used for rigidly rotating star in two ways.
    First, the full moment of inertia $I_{tot}$ in the denominator is replaced by the combinations of $|I_{crust}-S_2 I_{core}|$ and $S_3 I_{core}$ which are much less than $I_{tot}$. It leads to more rapid evolution of the angles. 
    Second,     
    one can see that Eq. \eqref{eq:chi_eq} contains $\tilde{k}_\perp$ (multiplied by $S_3$) which 
    does not influence the evolution of the inclination angle in the case of a rigid star (cf. \cite{CasiniMontemayor1998}).
    The precession is affected by $\tilde{k}_m$ through non-zero $S_3$ as well.
    
    The perpendicular part of $\vec{K}$ is associated with the so-called anomalous torque caused by inertia of the near-zone electromagnetic field \cite{Melatos2000}. 
    The anomalous torque is larger than the other components of $\vec{K}$ approximately by a  factor
    \begin{equation}
      \label{eq:anomalous_troque_factor}
      \frac{c}{\Omega R_{ns}} = 4.7\times10^3 \left(\frac{P}{1\mbox{sec}}\right)\left(\frac{R_{ns}}{10\mbox{km}}\right)^{-1}.
    \end{equation}    
    which grows as the star slows down under the action of the braking torque. As we show below, $S_2$ and $S_3$ also grow as the star evolves. Therefore, the influence of anomalous torque on the inclination angle evolution increases with time.     
    The effect becomes important when $S_3 \tilde{k}_\perp \approx \left(I_{crust}/I_{core}-S_2\right)\tilde{k}_m$.

    \section{Weak coupling limit}
    Let us first consider the case of magnetic field confined to the crust. In this case, crust-core interaction occurs through the viscosity.       
    If neutrons are not superfluid, all constituents can be regarded as moving with the same velocity $\vec{u}$.
    It can be shown that the composition gradient $\vec\nabla(\rho_p/\rho_n)$ causes a strong damping of any stationary radial flow. Thus, the angular momentum can be transfered from the core to the crust only by viscous tensions. In this case,  one finds that (detailed calculations can be found in \cite{BGT2014})
    \begin{equation}
       \label{eq:coefs_visc}
       S_2 = -\frac{8\pi \rho^{(b)} R_c^5}{3\sqrt{2}I_{core}}E^{1/2}, \ \ S_3 =  \frac{8\pi \rho^{(b)} R_c^5}{3\sqrt{2}I_{core}}E^{1/2},
    \end{equation}
    where $\rho^{(b)}$ is the mass density at the crust-core interface, $E = \eta^{(b)}/\Omega R_c^2$ is the so-called Ekman number, $\eta^{(b)}$ is the value of the shear viscosity coefficient at the crust-core interface. The expressions for these coefficients were obtained under the assumption that $E\ll 1$. It is satisfied with a good accuracy for neutron stars. However, $\eta$ essentially depends on temperature. Therefore, in order to close the system of equations one needs to take into account the  thermal evolution of the star, which  practically 
    does not depend on the spin evolution. It is determined mostly by the mass of the star, which
    implies that the viscosity is a  known function of star's age.
    
    If neutrons are superfluid, the neutron component decouples from the rest of the matter and an additional dynamical degree of freedom - the velocity of superfluid neutrons $\vec{u}_s$ - appears. The neutron fluid 
    and the charged components interact only by weak mutual friction force.
    In this situation, the damping of radial flows may not take place. Besides the viscous tensions, the angular momentum can be transfered directly by flow (the so-called Ekman pumping mechanism \cite{Greenspan_book}). 
    However, neutron superfluidity starts to affect the crust-core interaction only when the distance between the crust-core interface and phase transition surface is of the order of $E^{1/2}R_c$ or smaller. The realization of such a configuration depends on the model of neutron superfluidity.

  \section{Strong coupling limit}
    The neutron superfluidity becomes more important in the case of strong coupling. In this limit we  assume that the charged component and the normal neutrons rotate rigidly with the crust.  The second component which rotates differentially consists of superfluid neutrons. The components interact through the mutual friction. These crust-core interaction coefficients are given by (see \cite{BGT2013b} for the detail)
    \begin{equation}
      \label{eq:coefs_mf}
      S_2 = -\frac{I_c}{I_{core}}-\frac{8\pi}{3I_{core}}\int_0^{R_s}   \frac{(1-\beta')\beta'-\beta^2}{(1-\beta')^2+\beta^2}\rho_s r^4 dr, \ \
      S_3 = \frac{8\pi}{3I_{core}}\int_0^{R_{s}}\frac{\beta}{(1-\beta')^2+\beta^2}\rho_s r^4 dr,
    \end{equation}
    where $\beta$ and $\beta'$ are the mutual friction coefficients, $\rho_s$ is the mass density of superfluid neutrons, $R_s$ is the radius of the spherical surface of the phase transition.
    
    Note that during star's cooling more and more neutrons become superfluid, i.e., the normal neutron component, which co-rotates with the charge component, is converted into the superfluid component. It is not obvious that Eqs. \eqref{eq:omega_eq}-\eqref{eq:phi_eq} remain valid in this situation. The left-hand side of Eq. \eqref{eq:crust_eq} should also contain $\dot{I}_{crust}\vec{\Omega}$. Now the angular momentum exchange between components is produced not only by $\dot{\vec{\Omega}}$ but also directly by matter conversion. It can be shown that in the linear in $\epsilon$ approximation two terms should be added to the interaction torque \eqref{eq:N_torque}:
    \begin{equation}
      \vec{N}_{new} = \vec{N}_{old} - \dot{I}_{core} \vec{\Omega} - \int \dot{\rho}_s^{(0)} [\vec{r}\times\vec{u}_s] dV.
    \end{equation}
    The first additional term cancels with $\dot{I}_{crust} \vec{\Omega}$. As for the second term, we assume that the thermal evolution occurs on approximately the same time-scale as the rotational evolution. In that case, the second term is quadratically small and can be neglected.

  \section{Inclination angle evolution}
\begin{figure}
      \centering
      \includegraphics[width=0.42\textwidth]{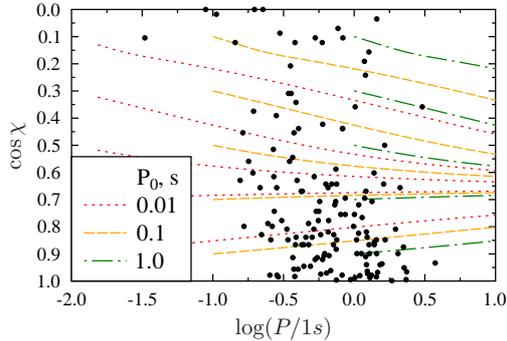}
      \caption{The $\cos\chi-P$ trajectories for rigidly rotating neutron stars with different initial periods and inclination angles.}
      \label{fig:BPT_traj}
    \end{figure}

    In order to plot some exemplary inclination angle trajectories we use the model expression for external torque proposed in \cite{BarsukovPolyakovaTsygan2009}  with non-dipolarity parameter $\nu_{nd}=0.5$. The trajectories for a rigidly rotating neutron star are shown in Fig. \ref{fig:BPT_traj}.
    
\begin{figure}
      \includegraphics[trim = 10 10 10 0, clip ,width=1.00\textwidth]{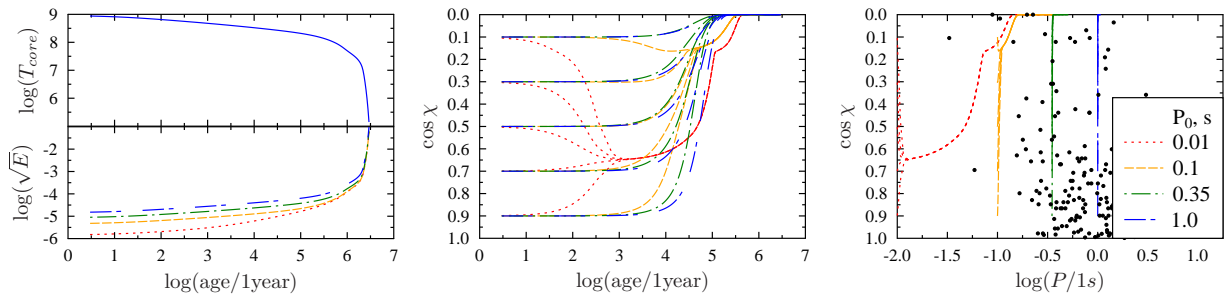} 
      \caption{The evolution of neutron star core temperature and $E^{1/2}$ (left panel), the inclination angle evolution trajectories (middle panel) and the same trajectories in $\cos\chi-P$ plane (right panel).}
      \label{fig:traj_BPT_I01_nu05}
    \end{figure}

    The trajectories calculated with the coefficients \eqref{eq:coefs_visc} are shown in Fig. \ref{fig:traj_BPT_I01_nu05}.
    Here, in order to take into account the thermal evolution of the star we have used the cooling code developed by Gnedin, Yakovlev and Potekhin \cite{GnedinYakovlevPotekhin2001}. The cooling curve has been calculated for a light star with a mass $\approx 1 M_\odot$ and $I_{crust}/I_{tot}$=0.05. 
    
\begin{figure}
	  \includegraphics[trim = 3 0 0 0, clip ,height=0.25\textwidth]{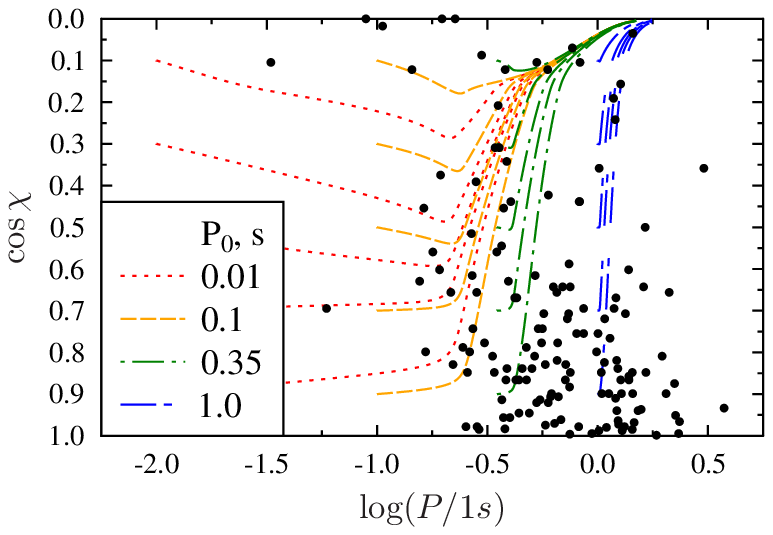}  
	  \includegraphics[height=0.25\textwidth]{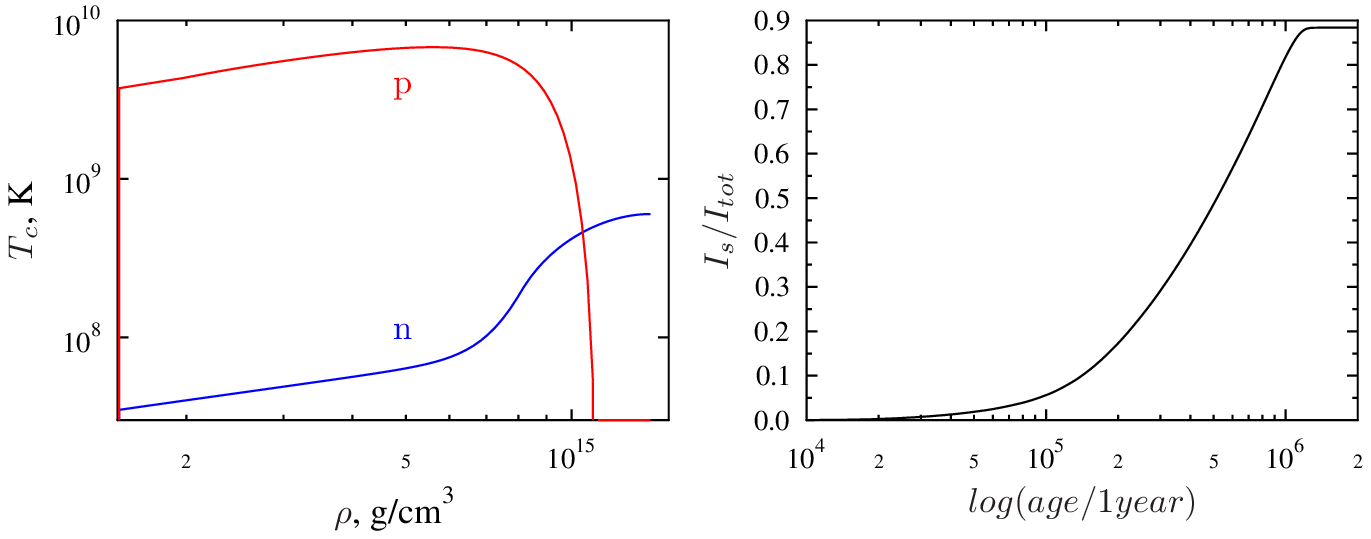}
	  \caption{The $P-\cos\chi$ trajectories (left panel), the models of proton and neutron superfluidities (middle panel) and the evolution of relation $I_{s}/I_{tot}$ with time (right panel).}
	  \label{fig:traj_BPT_sf1}
    \end{figure}

\begin{figure}
	  \includegraphics[trim = 3 0 0 0, clip ,height=0.25\textwidth]{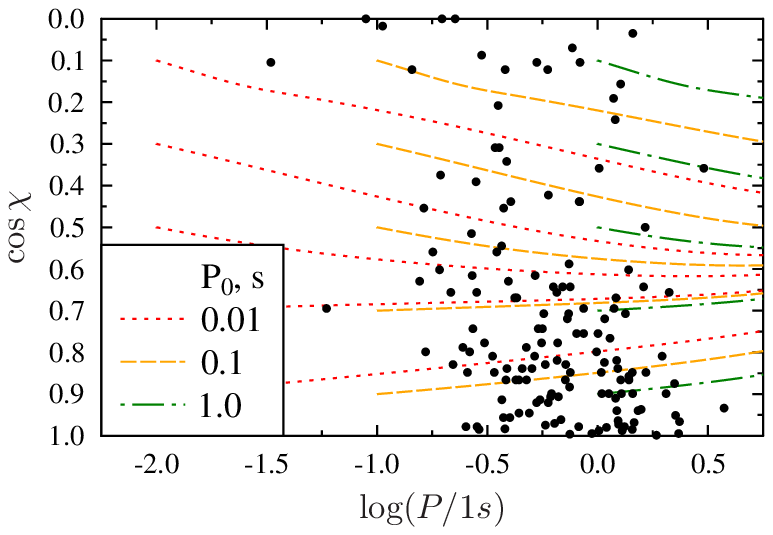}  
	  \includegraphics[height=0.25\textwidth]{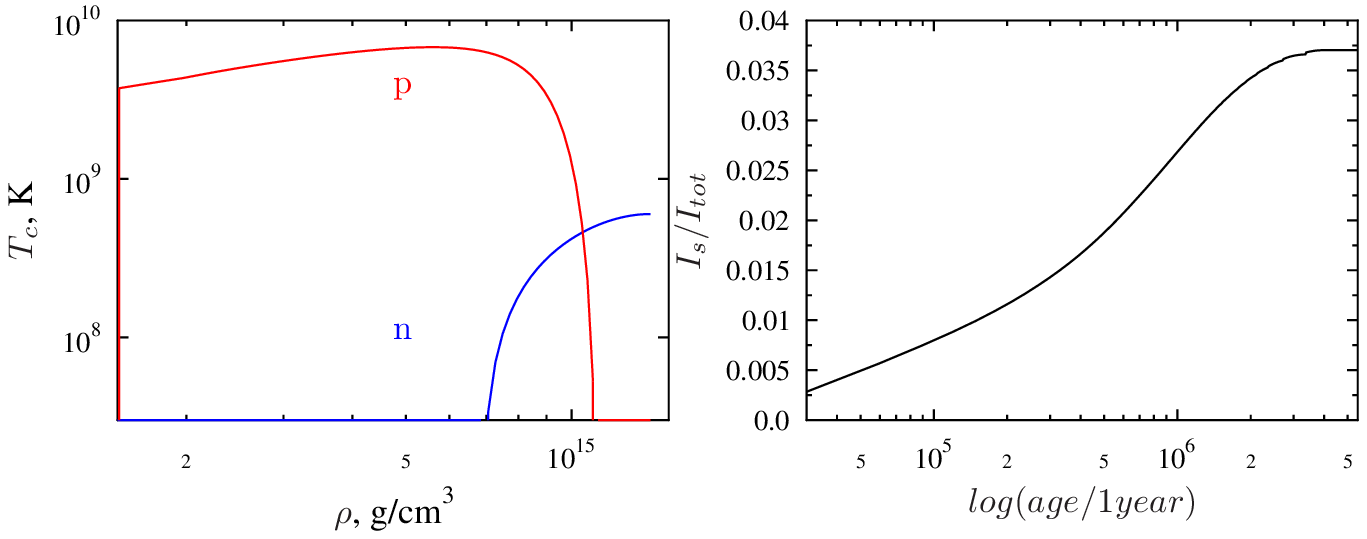}
	  \caption{The same as in Fig. \ref{fig:traj_BPT_sf1} for another model of neutron superfluidity.}
	  \label{fig:traj_BPT_sf2}
    \end{figure}

    The trajectories calculated with coefficients \eqref{eq:coefs_mf} for two different neutron superfluid models are shown in Figs. \ref{fig:traj_BPT_sf1} and \ref{fig:traj_BPT_sf2}. The superfluidity models used in Fig. \eqref{fig:traj_BPT_sf1} are taken from \cite{GusakovKaminkerYakovlevGnedin2005}. Fig. \ref{fig:traj_BPT_sf2} is plotted for a model in which the neutron superfluidity is strongly damped in the region of small densities.
    The mutual friction coefficients have been calculated using the results of Alpar, Langer and Sauls \cite{AlparLangerSauls1984}.
   All figures include 149 pulsars whose observable values of 
the inclination angle are taken from \cite{Rankin1993}.
    
  \section{Discussion}  

  One can see that the evolution of the inclination angle dramatically differs for different crust-core interaction models. The rate of the evolution depends on the amount of the matter which rotates differentially. Moreover, all trajectories sharply turn upwards when the action of anomalous torque becomes comparable with the action of normal torque. These features should be included in the studies of population synthesis of pulsars. They 
can also be used as additional tests for the theories of neutron star interiors.

     \section*{Acknowledgements}
    This work was supported by the Russian Foundation for the Basic Research (project 13-02-00112), the Programme of the State Support for Leading Scientific Schools of the Russian Federation (grant NSh-4035.2012.2) and Ministry of Education and Science of Russian Federation (Agreement No. 8409). 

  \section*{References}
  \bibliography{mn-jour,paper}

\providecommand{\newblock}{}
\begin{thebibliography}{10}
\expandafter\ifx\csname url\endcsname\relax
  \def\url#1{{\tt #1}}\fi
\expandafter\ifx\csname urlprefix\endcsname\relax\def\urlprefix{URL }\fi
\providecommand{\eprint}[2][]{\url{#2}}

\bibitem{Spitkovsky2008}
{Spitkovsky} A 2008 {\em 40 Years of Pulsars: Millisecond Pulsars, Magnetars
  and More\/} ({\em American Institute of Physics Conference Series\/} vol 983)
  pp 20--28

\bibitem{Easson1979}
{Easson} I 1979 {\em ApJ\/} {\bf 233} 711--716

\bibitem{YakovlevLevenfishShibanov1999}
Yakovlev D~G, Levenfish K~P and Shibanov Y~A 1999 {\em Physics-Uspekhi\/} {\bf
  42} 737

\bibitem{HallVinen1956}
{Hall} H~E and {Vinen} W~F 1956 {\em Royal Society of London Proceedings Series
  A\/} {\bf 238} 215--234

\bibitem{PonsMirallesGeppert2009}
{Pons} J~A, {Miralles} J~A and {Geppert} U 2009 {\em A\&A\/} {\bf 496} 207--216

\bibitem{GourgouliatosEtAl2013}
{Gourgouliatos} K~N {\em et~al.\/} 2013 {\em MNRAS\/} {\bf 434} 2480--2490

\bibitem{BuckleyMetlitskiZhitnitsky2004a}
{Buckley} K~B, {Metlitski} M~A and {Zhitnitsky} A~R 2004 {\em Physical Review
  Letters\/} {\bf 92} 151102

\bibitem{Link2006}
{Link} B 2006 {\em A\&A\/} {\bf 458} 881--884

\bibitem{GnedinYakovlevPotekhin2001}
{Gnedin} O~Y, {Yakovlev} D~G and {Potekhin} A~Y 2001 {\em MNRAS\/} {\bf 324}
  725--736

\bibitem{CasiniMontemayor1998}
{Casini} H and {Montemayor} R 1998 {\em ApJ\/} {\bf 503} 374

\bibitem{Melatos2000}
{Melatos} A 2000 {\em MNRAS\/} {\bf 313} 217--228

\bibitem{BGT2014}
{Barsukov} D~P, {Goglichidze} O~A and {Tsygan} A~I 2014 {in preparation}

\bibitem{Greenspan_book}
Greenspan H 1990 {\em The theory of rotating fluids\/} Cambridge monographs on
  mechanics and applied mathematics (At the University Press) ISBN
  9780962699801

\bibitem{BGT2013b}
{Barsukov} D~P, {Goglichidze} O~A and {Tsygan} A~I 2013 {\em MNRAS\/} {\bf 432}
  520--529

\bibitem{BarsukovPolyakovaTsygan2009}
{Barsukov} D~P, {Polyakova} P~I and {Tsygan} A~I 2009 {\em Astronomy Reports\/}
  {\bf 53} 1146--1154

\bibitem{GusakovKaminkerYakovlevGnedin2005}
{Gusakov} M~E, {Kaminker} A~D, {Yakovlev} D~G and {Gnedin} O~Y 2005 {\em
  MNRAS\/} {\bf 363} 555--562

\bibitem{AlparLangerSauls1984}
{Alpar} M~A, {Langer} S~A and {Sauls} J~A 1984 {\em ApJ\/} {\bf 282} 533--541

\bibitem{Rankin1993}
{Rankin} J~M 1993 {\em ApJS\/} {\bf 85} 145--161

\end{thebibliography}
    
\end{document}